\documentclass[aps,prd,twocolumn,superscriptaddress,nofootinbib,floatfix,noshowpacs]{revtex4-1}
\pdfoutput=1
\newif\ifcomment
\usepackage[T1]{fontenc}
\usepackage{amsmath,amssymb}
\usepackage{color,hyperref,url}
\usepackage{listings}
\usepackage{slashed}
\usepackage[pdftex]{graphicx}
\usepackage{epstopdf}
\usepackage{epsfig}
\usepackage{grffile}
\usepackage{relsize}
\graphicspath{{./img/}}
\usepackage{soul}
\usepackage{multirow}
\usepackage{adjustbox}

\newcommand{\beq}{\begin{equation}}
\newcommand{\eeq}{\end{equation}}
\newcommand{\ba}{\begin{array}}
\newcommand{\ea}{\end{array}}
\newcommand{\bea}{\begin{align}}
\newcommand{\eea}{\end{align}}
\newcommand{\bi}{\begin{itemize}}
\newcommand{\ei}{\end{itemize}}
\newcommand{\ben}{\begin{enumerate}}
\newcommand{\een}{\end{enumerate}}
\newcommand{\bc}{\begin{center}}
\newcommand{\ec}{\end{center}}
\newcommand{\bl}{\begin{flushleft}}
\newcommand{\el}{\end{flushleft}}
\newcommand{\br}{\begin{flushright}}
\newcommand{\er}{\end{flushright}}

\newcommand{\nn}{\nonumber \\}
\newcommand\Eqn[1]{Eq.~(\ref{#1})}  

\newcommand{\GeV}{{\rm GeV}}

\DeclareUnicodeCharacter{2212}{\ensuremath{-}}
\begin{document}
\title{A Direct Algebraic Pathway to Hadronic Observables in the Contact Model}

\author{Jiayin Kang}
\affiliation{School of Physics, Nankai University, Tianjin 300071, China}

\author{Zanbin Xing}\email{xingzb@mail.nankai.edu.cn}
\affiliation{School of Physics, Nankai University, Tianjin 300071, China}

\author{Lei Chang}\email{leichang@nankai.edu.cn}
\affiliation{School of Physics, Nankai University, Tianjin 300071, China}

\date{\today}

\begin{abstract}
We present a novel algebraic framework for computing hadron properties directly within the contact interaction model. Utilizing Fierz transformations, the method recasts the Bethe-Salpeter dynamics into equations for a minimal set of \emph{projected amplitudes} for bound-state static properties and form factors, bypassing the conventional need for the meson wave function. This approach is fully demonstrated for the vector meson, enabling the direct extraction of its decay constants and form factors. The formalism provides a more efficient and unified pathway to hadron observables, with clear potential for extension to baryons and more sophisticated interactions.
\end{abstract}


\maketitle

\section{Introduction}
Understanding the structure and properties of hadrons from the first principles of Quantum Chromodynamics (QCD) remains a central challenge in nuclear and particle physics~\cite{Gross:2022hyw}. In the non-perturbative regime, frameworks such as that of the  Dyson-Schwinger equations (DSEs)~\cite{Roberts:1994dr,Alkofer:2000wg,Bashir:2012fs,Fischer:2018sdj,Roberts:2021nhw,Eichmann:2025wgs} have proven fruitful. The contact interaction (CI) model, first systematically introduced in the DSEs context in Ref.~\cite{Gutierrez-Guerrero:2010waf} and subsequently developed in Refs.~\cite{Roberts:2010rn,Roberts:2011cf,Roberts:2011wy,Xu:2015kta,Raya:2017ggu,Serna:2017nlr,Gutierrez-Guerrero:2019uwa,Yin:2019bxe,Zhang:2020ecj,Xu:2021iwv,Xing:2022sor,Xing:2021dwe,Wang:2022mrh,Xing:2022jtt,Xing:2022mvk,Xing:2023eed,Dang:2023ysl,Paredes-Torres:2024mnz,Hernandez-Pinto:2024kwg,Xing:2024bpj,Sultan:2024hep,Xu:2024frc}, serving as a tractable infrared approximation, is widely adopted within DSEs studies both for pedagogical purposes and for quantitative hadronic calculations using a symmetry-preserving regularization scheme~\cite{Xing:2022jtt}.

The conventional paradigm in DSEs studies follows a well-established sequence~\cite{Llewellyn-Smith:1969bcu,Munczek:1991jb,Roberts:1994dr,Tandy:1997qf,Hernandez-Pinto:2024kwg}: a symmetry-preserving truncation, such as the rainbow-ladder (RL) approximation, is first employed to derive the quark propagator; subsequently, the bound-state Bethe-Salpeter or Faddeev equations are solved to obtain wave functions; finally, these wave functions are used to compute observable quantities like decay constants and form factors within the symmetry preserving framework. 

Although robust, this multi-step procedure requires the intermediate determination of wave functions, which are not themselves direct physical observables. In this work, we propose and demonstrate an alternative, streamlined paradigm that circumvents this conventional sequence. By making judicious use of Fierz transformations, we derive closed algebraic equations that directly govern the projected amplitudes for physical quantities—bypassing the explicit calculation of bound-state wave functions altogether. This approach offers a more efficient and conceptually direct pathway from the model interaction to measurable hadronic properties.

To illustrate the efficacy and simplicity of this new scheme, we present a detailed calculation for the vector meson as a benchmark case~\cite{Xing:2021dwe,Hernandez-Pinto:2024kwg}. The procedure, however, is general and can be readily applied to other meson systems as well as baryons. Beyond computational efficiency, this algebraic reformulation offers a novel and unified perspective on hadron structure, directly linking model interactions to a broad class of observables without intermediary wave functions.

The remainder of this paper is organized as follows. 
In Sec.~\ref{sec:equations_observables}, we present the general formalism: starting from the Fierz-transformed contact interaction, we derive the homogeneous and inhomogeneous algebraic equations that directly govern the projected amplitudes for bound-state properties and form factors, respectively. 
Section~\ref{sec:rho_observables} applies this formalism specifically to the vector meson, detailing the tensor structures of its projected amplitudes and the complete basis of form factors. 
The corresponding numerical results are also presented and discussed in this section. 
Finally, Sec.~\ref{sec:summary} summarizes our findings and outlines future perspectives. 
Useful algebraic details are compiled in the Appendices.

\section{Equations for Physical Observables}
\label{sec:equations_observables}
In this section, we derive the closed-form equations satisfied directly by the projected amplitudes of physical observables within the CI model. This is achieved by applying Fierz transformations to the standard Bethe-Salpeter equation(BSE), followed by appropriate Dirac projections, thereby eliminating the need to first solve for the full Bethe-Salpeter amplitudes (BSAs).

The starting point is the Fierz reordering identity:
\begin{equation}
   \gamma_{\alpha}^{ij}\gamma_{\alpha}^{kl} - \xi\,\tilde{\Gamma}_j^{ij}\tilde{\Gamma}_j^{kl} = \mathcal{C}_{aa'}\, \Lambda^{a}_{il}\Lambda^{a'}_{kj},
   \label{eq:fierz_identity} 
\end{equation}
where the second interaction channel (represents the anomalous magnetic moment (AMM) effect\cite{Xing:2021dwe}) is defined by $\tilde{\Gamma}_j = \{I_4, \gamma_5, i\sigma_{\alpha\beta}/\sqrt{6}\}$ with a strength parameter $\xi$. The coefficient matrix is $\mathcal{C}_{aa'} = \mathrm{diag}\{1, -1, -1/2, -1/2, -\xi/3\}$, and the Dirac basis is $\{\Lambda^{a}\}=\Lambda^{\{S,P,V,A,T\}}_{(\mu_a \nu_a)} = \{I, \gamma_{5}, \gamma_{\mu_a}, \gamma_{5}\gamma_{\mu_a}, \sigma_{\mu_a\nu_a}\}$. Herein we use the shorthand $a,\,b,\,c$ to denote indices running over the set $\{S,P,V,A,T\}$ and suppress the possible Lorentz indices for brevity. Repeated indices imply summation.

\subsection{Homogeneous Equation: Masses and Decay Constants}
\label{sub:homogeneous}

To derive the closed-form equations, we start from the meson BSE in the CI model:
\begin{align}
    \Gamma_H(Q) = &-D_G \int_q \gamma_\alpha \, \chi_H(q,Q) \, \gamma_\alpha \nn
    & -D_G \xi \int_q \tilde{\Gamma}_j \, \chi_H(q,Q) \, \tilde{\Gamma}_j\,,
\end{align}
where $\int_q=\int\frac{d^4q}{(2\pi)^4}$, $D_G = 4/(3 m_G^2)$, $\Gamma_H(Q)$ is the Bethe-Salpeter amplitude of meson type $H$ and the corresponding Bethe-Salpeter wave function is $\chi_H(q,Q) = S(q_+) \Gamma_H(Q) S(q_-)$ with $X_\pm = X \pm Q/2$. The quark propagator satisfies the corresponding DSE \footnote{The quark propagator is actually a diagonal matrix in flavour space denoted by $\mathcal{S}=\mathrm{diag}\{S_u,S_d,...\}$. For brevity we focus on the two lightest quarks and work in the isospin limit: $S_u(p)=S_d(p)\equiv S(p)$.}
\begin{equation}
    S^{-1}(p)=i\slashed{p}+m+D_G \int_q \gamma_\alpha \, S(q) \, \gamma_\alpha  \,,
\end{equation}
and it can be parameterized as $S^{-1}(p)=i\slashed{p}+M$, we apply the symmetry preserving regularization scheme proposed in \cite{Xing:2022jtt} and relevant parameters therein, which are presented in Table \ref{tab:static}.

Applying the Fierz identity \Eqn{eq:fierz_identity} transforms the BSE into an algebraic form:
\begin{equation}\label{eq:BSA_algebraic}
    \Gamma_H(Q) = \Lambda_{a} \, \mathcal{D}_{aa'} \, g^H_{a'}(Q),
\end{equation}
where $\mathcal{D}_{aa'} \equiv -D_G\,\mathcal{C}_{aa'}$. Here, we have introduced the key projected amplitudes $g^H_a(Q)$, defined as
\begin{equation}\label{eq:def_ga}
    g^H_a(Q) = \mathrm{tr} \int_q \Lambda_a \, \chi_H(q,Q),
\end{equation}
with tr denoting the Dirac trace. These amplitudes carry the remaining color and flavor indices but encapsulate the essential dynamical information. \Eqn{eq:BSA_algebraic} shows that the full BSA in the CI model is completely determined by a finite set of these projected amplitudes\footnote{For the general momentum-dependent interaction model the  number of projected amplitudes is infinite.}.

To obtain an equation governing $g^H_a(Q)$, we multiply both sides of the original BSE by the projection operator $S(p_-) \Lambda_b S(p_+)$, take the Dirac trace, and integrate over the relative quark momentum $p$. This procedure yields:
\begin{equation}\label{eq:g_eigen}
    g^H_{b}(Q) = \Pi_{ba}(Q) \, \mathcal{D}_{aa'} \, g^H_{a'}(Q),
\end{equation}
where the kernel is
\begin{equation}
    \Pi_{ba}(Q) = \mathrm{tr} \int_p  S(p_-) \Lambda_b S(p_+) \Lambda_a.
\end{equation}
\Eqn{eq:g_eigen} is a homogeneous equation for the projected amplitudes. It can be recast as an eigenvalue problem,
\begin{equation}
    \lambda(Q^2) \, g^H_{b}(Q) = \Pi_{ba}(Q) \, \mathcal{D}_{aa'} \, g^H_{a'}(Q),
\end{equation}
with the physical bound-state mass $m_H$ identified by the condition $\lambda(Q^2 = -m_H^2) = 1$.

The canonical normalization condition for the BSA can also be expressed solely in terms of $g^H_a(Q)$. Starting from the standard normalization condition ~\cite{Llewellyn-Smith:1969bcu,Nakanishi:1969ph}, and using \Eqn{eq:BSA_algebraic}, one finds:
\begin{align}\label{eq:norm_ga}
    1 &= \left. \frac{\partial \ln \lambda(Q^2)}{\partial Q^2} \, \mathcal{D}_{aa'} \, \mathrm{Tr}\left[ \, \bar{g}^H_a(-Q) \, g^H_{a'}(Q) \right] \right|_{Q^2 = -m_H^2},\nn
    &=\left. \frac{\partial \ln \lambda(Q^2)}{\partial Q^2} \, \mathrm{Tr}[ \bar{g}^H_a(-Q)\,\Pi^{-1}_{ab}(Q)\,g^H_{b}(Q)\right|_{Q^2 = -m_H^2}
\end{align}
where $\mathrm{Tr}$ denotes color and flavor traces, and $\bar{g}^H_a(-Q)$ is the charge-conjugate amplitude, related to $g^H_a(Q)$ via $\bar{\chi}_H(q,-Q) = C \chi_H^{\mathsf{T}}(-q,-Q) C^{-1}$ in \Eqn{eq:def_ga}. The first row is a simple formula that could be applied in CI model, while the second row presents the normalization in hadronic level since all partonic information are integrated out and is not limited for CI.

\subsection{Inhomogeneous Equation: Form Factors and Transition Amplitudes}
\label{sub:inhomogeneous}

For processes involving an external current, we consider the generalized quark-quark correlation $\mathcal{M}_H(q,P,Q)$, where $P=(P_i+P_f)/2$ and $Q=P_f-P_i$ are the average and transfered momentum, satisfying $P^2 = -m_H^2 - Q^2/4$ and $P\cdot Q=0$ for on-shell mesons. Its amputated version $F(q,P,Q)$ satisfies an inhomogeneous BSE~\cite{Xing:2022mvk,Cotanch:2002vj}:
\begin{align}
    F_H(p,P,Q) &= F^{ih}_H(p,P,Q) - D_G \int_q \gamma_\alpha \, \mathcal{M}_H(q,P,Q) \, \gamma_\alpha \nn
             & - D_G \xi \int_q \tilde{\Gamma}_j \, \mathcal{M}_H(q,P,Q) \, \tilde{\Gamma}_j,
\end{align}
with $\mathcal{M}_H(q,P,Q)=S(q_-)F_H(q,P,Q)S(q_+)$, and the inhomogeneous term $F^{ih}_H(p,P,Q) = \Gamma_H(P_i) S(p-P) \bar{\Gamma}_H(-P_f)$.

Following the same Fierz transformation and projection technique, we obtain:
\begin{equation}
    F_H(p,P,Q) = F^{ih}_H(p,P,Q) + \Lambda_a \, \mathcal{D}_{aa'} \, \mathcal{A}^H_{a'}(P,Q),
\end{equation}
where the {\it invariant amplitudes} $\mathcal{A}^H_a(P,Q)$ are defined as
\begin{equation}\label{eq:def_Aa}
    \mathcal{A}^H_a(P,Q) = \mathrm{tr} \int_q \Lambda_a \, \mathcal{M}_H(q,P,Q).
\end{equation}
Projecting with $S(p_+)\Lambda_c S(p_-)$ leads to the governing equation for these amplitudes:
\begin{equation}\label{eq:A_inhomogeneous}
    \mathcal{A}^H_c(P,Q) = \mathcal{A}^{H,ih}_c(P,Q) + \Pi_{ca}(-Q) \, \mathcal{D}_{aa'} \, \mathcal{A}^H_{a'}(P,Q).
\end{equation}
The inhomogeneous term $\mathcal{A}^{H,ih}_c(P,Q)$ is computed from $F^{ih}_H$ and can be expressed by using the previously defined amplitudes $g^H_a$:
\begin{equation}
    \mathcal{A}^{H,ih}_c(P,Q) = \mathcal{D}_{aa'} \mathcal{D}_{bb'} \, g^H_{a'}(P_i) \, \bar{g}^H_{b'}(-P_f) \, \mathcal{T}_{cab}(P,Q),
\end{equation}
where the universal triangle diagram function is
\begin{equation}
    \mathcal{T}_{cab}(P,Q) = \mathrm{tr} \int_p \Lambda_c S(p_-) \Lambda_a S(p-P) \Lambda_b S(p_+).
\end{equation}
We note that in Ref.~\cite{Xing:2022mvk}, the treatment of the contact interaction inevitably introduced self-energy terms into the scattering amplitude. Our present approach avoids this complication, rendering the calculation more direct and generalizable.

\subsection{Structure and Universality of the projected amplitudes}
\label{sub:summary_obs}

The formalism developed above exhibits a clear, factorized structure:
\begin{itemize}
    \item Two Types of Basis Amplitudes: The dynamics of a meson $H$ are encoded in two sets of projected amplitudes:
    \begin{align}
        \chi_H(q,Q) \quad &\longleftrightarrow \quad \{g^H_{a}(Q)\}, \\
        \mathcal{M}_H(q,P,Q) \quad &\longleftrightarrow \quad \{\mathcal{A}^H_{a}(P,Q)\},
    \end{align}
    governed by the homogeneous \Eqn{eq:g_eigen} and inhomogeneous \Eqn{eq:A_inhomogeneous}, respectively. The normalization condition \Eqn{eq:norm_ga} and the inhomogeneous term in \Eqn{eq:A_inhomogeneous} depend solely on $g^H_a$.

\item Universal Kernels: The functions $\Pi_{ba}(Q)$ and $\mathcal{T}_{cab}(P,Q)$ are {\it universal}; they depend only on the quark propagator and the Dirac basis, not on the specific meson quantum numbers. All meson-specific information is coded by the corresponding quantum numbers and contained within the color-flavor structure of $g^H_a$ and $\mathcal{A}^H_a$.

\item Basis of Observables: The sets $\{g^H_a\}$ and $\{\mathcal{A}^H_a\}$ for a given $\xi$ form a basis. Once the basis observables are known, the projected amplitudes $g^H_\Gamma$ or $\mathcal{A}^H_\Gamma$ with arbitrary project operator $\Gamma$ instead of the basis operators $\Lambda$ can be expressed in terms of these basis amplitudes:
\begin{equation}
    g^H_{\Gamma}(Q) = \Pi_{\Gamma a}(Q) \, \mathcal{D}_{aa'} \, g^H_{a'}(Q),
\end{equation}
\begin{equation}
    \qquad\mathcal{A}^H_\Gamma(P,Q) = \mathcal{A}^{H,ih}_\Gamma(P,Q) + \Pi_{\Gamma a}(-Q) \, \mathcal{D}_{aa'} \, \mathcal{A}^H_{a'}(P,Q).
\end{equation}
For instance, in the rainbow-ladder (RL) truncation ($\xi=0$), the tensor ($a=T$) amplitudes are not basis amplitudes but can be expressed in terms of the vector ($a=V$) basis amplitudes. The gravitational form factors can also be obtained using a similar method.

\item Tensor Structure: After factoring out color and flavor (denoted by a tilde, e.g., $\tilde{g}^H_a$), the Lorentz-Dirac structure of these amplitudes is simple. For the homogeneous case:
    \begin{equation}
        \tilde{g}^H_a(Q) = \sum_i T^a_i(Q) \, \tilde{f}^H_{a,i},
    \end{equation}
    where $T^a_i(Q)$ are tensors formed from $Q$ and the metric/antisymmetric tensors, and $\tilde{f}^H_{a,i}$ are Lorentz-invariant coefficients. The structure of $\tilde{\mathcal{A}}^H_a(P,Q)$ is analogous to a meson-$a$-meson vertex:
    \begin{equation}
        \tilde{\mathcal{A}}^H_a(P,Q) = \sum_i T^a_i(P,Q) \, \tilde{F}^H_{a,i}(Q^2),
    \end{equation}
    where $T^a_i(P,Q)$ now depend on both $P$ and $Q$, and $\tilde{F}^H_{a,i}(Q^2)$ are the sought-after form factors.

\end{itemize}

\section{Vector Meson Observables}
\label{sec:rho_observables}
We proceed to apply the general formalism of Section II to vector mesons, a key class of hadrons characterized by spin-1 and odd parity. Focusing on the $\rho$ meson as a representative case, we detail the implementation of our approach: constructing the appropriate tensor structures, identifying the complete basis of form factors, and extracting physical observables such as decay constants, electromagnetic form factors. For convenience we suppress the explicit meson type in the projected amplitudes $g_a(P)$ and $\mathcal{A}_a(P,Q)$ in this section.

\subsection{Basis Projected Amplitudes}
\label{sub:basis_ff}

For a $\rho$ meson with momentum $P$ and polarization vector $\varepsilon$, the Lorentz-Dirac structures of the homogeneous projected amplitudes, $\tilde{g}_a(P)$ (obtained after factoring out color and flavour), are constrained by parity, charge conjugation, and angular momentum. They take the following form on the mass shell\footnote{We employ an Euclidean metric with $\{\gamma_\mu,\gamma_\nu\} = 2\delta_{\mu\nu}$; $\gamma_\mu^\dagger = \gamma_\mu$; $\gamma_5= \gamma_4\gamma_1\gamma_2\gamma_3$ so that $\text{tr}(\gamma_5\gamma_a\gamma_b\gamma_c\gamma_d)=-4\epsilon_{abcd}$; and $a \cdot b = \sum_{i}^{4} a_i b_i$.}:
\begin{align}
    \tilde{g}_V &= \varepsilon^\mu \, \tilde{f}_V, \\
    \tilde{g}_T &= \left(\varepsilon^{\mu} P^{\nu} - \varepsilon^{\nu} P^{\mu}\right) \tilde{f}_T,
\end{align}
where $\tilde{f}_V$ and $\tilde{f}_T$ are the independent Lorentz-invariant amplitudes to be determined from \Eqn{eq:g_eigen}. Substituting this structure into \Eqn{eq:g_eigen} reduces it to a coupled eigenvalue problem for the eigenvector $\{\tilde{f}_V, \tilde{f}_T\}$.

The corresponding structures for the invariant amplitudes $\tilde{\mathcal{A}}_a(P,Q)$ 
associated with the $\rho-a-\rho$ transition are more numerous. They are constructed from the available vectors $P_\mu$, $Q_\mu$, the polarization vectors $\varepsilon$, $\varepsilon'^*$, and the metric tensor. A complete basis can be chosen as:
\begin{align}
    \tilde{\mathcal{A}}_S &= s \, \tilde{F}_{S,1} + s_{PP} \, \tilde{F}_{S,2}, \\
    \tilde{\mathcal{A}}_P &= \epsilon_{PQ\varepsilon\varepsilon'^*} \, \tilde{F}_{P}, \\
    \tilde{\mathcal{A}}_V &= P_\mu \, s \, \tilde{F}_{V,1} + P_\mu \, s_{PP} \, \tilde{F}_{V,2} + 2 s_{\mu P} \, \tilde{F}_{V,3}, \\
    \tilde{\mathcal{A}}_A &= \epsilon_{\mu P \varepsilon\varepsilon'^*} \, \tilde{F}_{A,1} + Q_\mu \, \epsilon_{P Q \varepsilon\varepsilon'^*} \, \tilde{F}_{A,2} + 2 s_{\nu P} \, \epsilon_{\mu \nu P Q} \, \tilde{F}_{A,3}, \\
    \tilde{\mathcal{A}}_T &= s (P_\mu Q_\nu - Q_\mu P_\nu) \, \tilde{F}_{T,1} + s_{PP} (P_\mu Q_\nu - Q_\mu P_\nu) \, \tilde{F}_{T,2} \nn
    &\quad + 2 (Q_\mu s_{\nu P} - Q_\nu s_{\mu P}) \, \tilde{F}_{T,3} \nn 
    &\quad + 2 (P_\mu a_{\nu P} - P_\nu a_{\mu P}) \, \tilde{F}_{T,4} + 2 a_{\mu\nu} \, \tilde{F}_{T,5}.
\end{align}
Here, we have defined the following symmetric and antisymmetric tensors:
\begin{align}
    & s_{\mu\nu} = \frac{1}{2}(\varepsilon_\mu\varepsilon'^*_\nu + \varepsilon_\nu\varepsilon'^*_\mu), \quad 
    a_{\mu\nu} = \frac{1}{2}(\varepsilon_\mu\varepsilon'^*_\nu - \varepsilon_\nu\varepsilon'^*_\mu), \\
    & s= s_{\mu\mu}, 
    s_{PP} = s_{\mu\nu}P_\mu P_\nu, 
    s_{\mu P} = s_{\mu\nu} P_\nu,  
    a_{\mu P} = a_{\mu\nu} P_\nu.
\end{align}
The transverse projector is $\delta^T_{\mu\nu} = Q_\mu Q_\nu - \delta_{\mu\nu} Q^2$.

Upon substituting these structures into the inhomogeneous equation \Eqn{eq:A_inhomogeneous}, the system decouples into several subsets. Under non-zero $\xi$, The form factors $\tilde{F}_{S,1}$, $\tilde{F}_{S,2}$, $\tilde{F}_{A,1}$, $\tilde{F}_{A,3}$, $\tilde{F}_{T,4}$, and $\tilde{F}_{T,5}$ satisfy simple, uncoupled equations. 
In contrast, $\tilde{F}_{V,{i}}$ couples to $\tilde{F}_{T,{i}}$ for $i=1,2,3$, and $\tilde{F}_{P}$ couples to $\tilde{F}_{A,2}$, and then the axial form factor $\tilde{F}_{A,2}$ exhibits a pseudo scalar pole while the tensor form factors $\tilde{F}_{T,\{1,2,3\}}$ exhibits vector pole. Furthermore, the form factors $\tilde{F}_{T,{i+2}}$ and $\tilde{F}_{A,1}$ enter the respective coupled equations for $\{\tilde{F}_{V,{i}}, \tilde{F}_{T,{i}}\}$ and $\{\tilde{F}_P, \tilde{F}_{A,2}\}$ for $i=2,3$ as inhomogeneous driving terms.

The standard physical observables are obtained from these invariant amplitudes after including color and flavor factors ($N_c = 3$). The decay constant $f_\rho$ and its transverse counterpart $f_\rho^\perp$ are:
\begin{align}
    m_\rho f_\rho &= N_c \, \tilde{f}_V, \\
    m_\rho^2 f_\rho^\perp &= N_c \, \tilde{f}_T.
\end{align}
The dimensionless electromagnetic and transition form factors are conventionally defined as:
\begin{align}
    F_{S,1} &= N_c \, m_\rho^{-1} \, \tilde{F}_{S,1}, &
    F_{S,2} &= N_c \, m_\rho \, \tilde{F}_{S,2} /2, \\
    F_{P} &= - N_c \, m_\rho \, \tilde{F}_{P}, &
    F_{V,1} &= i N_c \, \tilde{F}_{V,1} / 2, \\
    F_{V,2} &= -i N_c \, m_\rho^{2} \, \tilde{F}_{V,2} / 4, &
    F_{V,3} &= -i N_c \, \tilde{F}_{V,3} / 2, \\
    F_{A,1} &= i N_c \, \tilde{F}_{A,1}, &
    F_{A,2} &= i N_c \, m_\rho^{2} \, \tilde{F}_{A,2}, \\
    F_{A,3} &= i N_c \, m_\rho^{2} \, \tilde{F}_{A,3}, &
    F_{T,1} &= i N_c \, m_\rho \, \tilde{F}_{T,1}, \\
    F_{T,2} &= i N_c \, m_\rho^{3} \, \tilde{F}_{T,2}, &
    F_{T,3} &= i N_c \, m_\rho \, \tilde{F}_{T,3} /2, \\
    F_{T,4} &= i N_c \, m_\rho \, \tilde{F}_{T,4}, &
    F_{T,5} &= i N_c \, m_\rho^{-1} \, \tilde{F}_{T,5}.
\end{align}
The electric-monopole, magnetic-dipole and electric-quadrupole form factors are associated with a set of irreducible tensors basis\cite{Arnold:1979cg}, their relations with the vector form factors defined herein are 
\begin{align}
    G_E(Q^2)&=F_1(Q^2)+\frac{2}{3}\frac{Q^2}{4m_\rho^2}G_Q(Q^2)\,,\nn
    G_M(Q^2)&=F_3(Q^2)\,,\nn
    G_Q(Q^2)&=F_1(Q^2)-F_3(Q^2)+F_2(Q^2)[1+\frac{Q^2}{4m_\rho^2}]\,.
\end{align}
The axial vector and tensor form factors could be similarly interpreted in analogy with the electromagnetic case.

\subsection{Numerical results}
\label{sbsec:numerical_results}

Given the scope of this paper, we provide only a broad overview of the calculations, without delving into technical details.

The static properties obtained from the homogeneous equation are listed in Table~\ref{tab:static}. It is noted that the parameter $\xi=0$ recovers original RL truncation, while the choice $\xi=0.151$ is fixed to reproduce the complete chiral anomaly under the present parameters \cite{Dang:2023ysl,Xing:2024bpj}. As shown in Table~\ref{tab:static}, the parameter $\xi$ also lowers the mass and enhances $f_\rho^\perp$, while $f_\rho$ remains nearly unchanged.

\begin{table}[htbp!]
\caption{\label{tab:static} constituent quark mass and static properties of the $\rho$ meson, mass units in GeV. In the CI framework, the chosen parameters are: $m_G=0.132\,\GeV$, $\tau_{ir}=1/0.24\,\GeV^{-1}$, $\tau_{uv}=1/0.905\,\GeV^{-1}$, $m=0.007\,\GeV$.}
\begin{tabular*}{0.45\textwidth}{c @{\extracolsep{\fill}} cccc}
\hline\hline
$\xi$ &$M$&$M_\rho$ &$f_\rho$ & $f_\rho^\perp$\\
\hline
0 &0.36769 &0.928672  &0.182747& 0.218621\\
0.151&0.36769 &0.879281 & 0.181862& 0.290463\\
\hline\hline
\end{tabular*}
\end{table}

\begin{figure}[htbp!]
    \centering
    \includegraphics[width=0.8\linewidth]{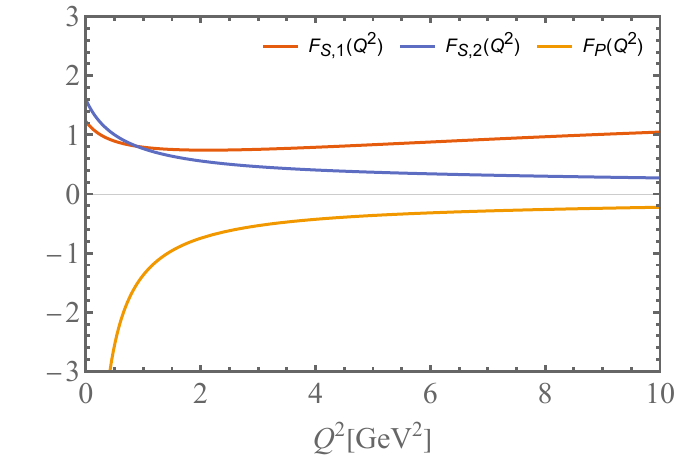}
    \caption{Scalar and pseudo scalar form factors. $F_{S,\{1,2\}}(Q^2)$ possess scalar meson pole while $F_{P}(Q^2)$ possesses pseudo scalar meson pole.}
    \label{fig:sff}
\end{figure}

\begin{figure}[htbp!]
    \centering
    \includegraphics[width=0.8\linewidth]{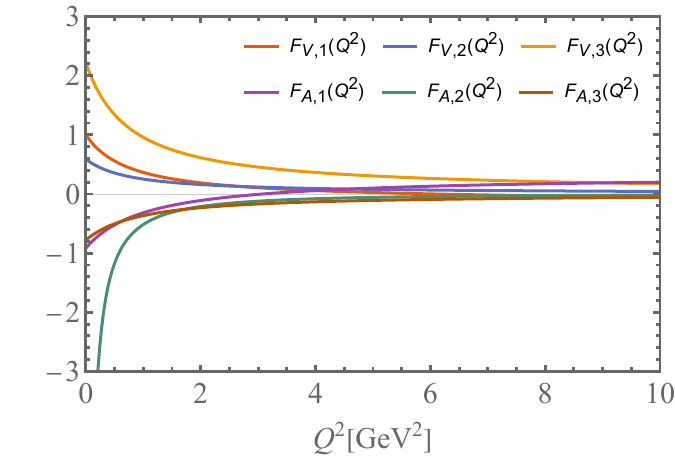}
    \caption{Vector and axial vector form factors. $F_{V,\{1,2,3\}}(Q^2)$ possess vector meson pole while $F_{A,\{1,2,3\}}(Q^2)$ possess axial vector meson pole. The form factor $F_{A,2}(Q^2)$ possesses an additional pseudo scalar meson pole.}
    \label{fig:vff}
\end{figure}

\begin{figure}[htbp!]
    \centering
    \includegraphics[width=0.8\linewidth]{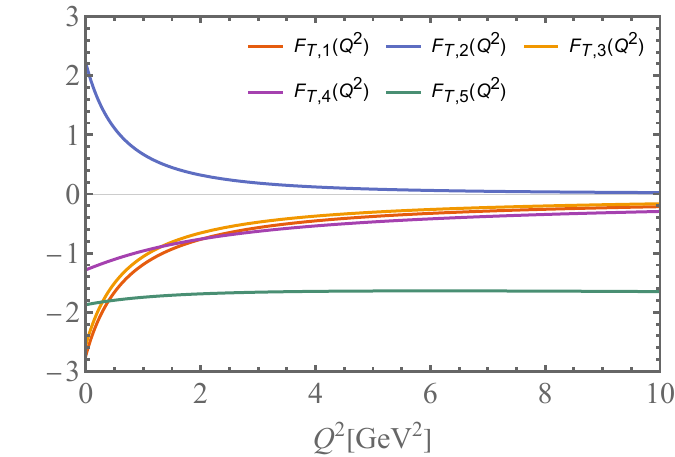}
    \caption{Tensor form factors. $F_{T,\{1,2,3\}}(Q^2)$ possess vector meson poles while $F_{T,\{4,5\}}(Q^2)$ do not possess any meson pole.} 
    \label{fig:tff}
\end{figure}

For the form factors, we choose to present them uniformly under $\xi=0.151$ because of the anomaly related pseudo scalar form factor and axial vector form factor.
The computed scalar, pseudoscalar, vector, and tensor form factors, which are obtained by solving the self-consistent, closed system of equations, are 
displayed in Fig.~\ref{fig:sff} (scalar and pseudoscalar), Fig.~\ref{fig:vff} (vector/electromagnetic and axial vector), and Fig.~\ref{fig:tff} (tensor), respectively. For clarity of the figure, we choose not to show the timelike region but instead describe the associate poles with corresponding form factors in the captions of figures, which could essentially affect the slope of the form factors at $Q^2=0$. While in the ultraviolet region, the form factors demonstrate relative stiffness, which are typical features of the CI model. Note that the scalar form factor $F_{S,1}(Q^2)$ exhibits non-monotonic behaviour, similar behaviour is also found in the pion scalar form factors.
These features—relative stiffness in the ultraviolet and meson poles in the timelike region—are characteristic of the contact interaction model and are consistent with earlier studies of vector mesons within similar frameworks~\cite{Xing:2021dwe, Hernandez-Pinto:2024kwg}. The behavior underscores the model’s effective description of low-momentum dynamics while retaining a simple analytic structure.

\section{summary}
\label{sec:summary}
In summary, we have developed and demonstrated a novel algebraic paradigm for computing hadronic observables directly within the framework of the contact interaction (CI) model and the Dyson-Schwinger equations. Moving beyond the conventional multi-step procedure---which requires sequential solutions for the quark propagator, the bound-state wave function, and finally the observables---we have systematically employed Fierz transformations and recasted the integral equations into a closed set of algebraic equations that govern a finite number of projected amplitudes, $g^H_a(P)$ and $\mathcal{A}^H_a(P,Q)$.

The core advantage of this method is that it completely bypasses the explicit calculation of the bound-state wave function, establishing a direct and efficient mapping from the model interaction to physical observables. Using the $\rho$ meson as a detailed case study, we have illustrated the full application of this paradigm: deriving the specific tensor structures of the projected amplitudes, identifying the complete set of basis form factors, and demonstrating the extraction of physical quantities such as decay constants ($f_\rho$, $f_\rho^\perp$), various form factors.

While demonstrated here for two-body meson systems, this approach is readily extensible to three-body bound-state equations, opening a promising route for the direct calculation of baryon properties—such as nucleon masses, form factors, and partonic distributions—within the same streamlined framework. A particularly worthwhile future direction is to establish more explicit connections between the projected amplitudes introduced here and quantities like parton distribution functions (PDFs) and distribution amplitudes (DAs), thereby unifying the description of elastic and inelastic hadron structure within a single, computationally efficient formulation.

\section{Acknowledgments} Work supported by: National Natural Science Foundation of China, grant no. 12135007; Postdoctoral Fellowship Program of CPSF under Grant Number GZC20240759.

\appendix

\section{$\Pi_{ba}(Q)$ and $\mathcal{T}_{cab}(P,Q)$ in CI}
When evaluating the function {$\Pi_{ba}(Q)$ and $\mathcal{T}_{cab}(P,Q)$ in CI, a proper regularization procedure is required. We use the symmetry preserving regularization scheme proposed in \cite{Xing:2022jtt}. The final expressions are expressed in terms of the so called irreducible loop integrals~\cite{Wu:2002xa}
\begin{equation}
    I_{-2\alpha}(s) :=\int_{q}\frac{1}{(q^{2}+s)^{\alpha+2}} \to \int_{\tau_{uv}^{2}}^{\tau_{ir}^{2}}d\tau\frac{\tau^{\alpha-1}}{\Gamma(\alpha+2)}\frac{e^{-\tau s}}{16\pi^{2}}
\end{equation}
The arrow denotes its form under proper time regularization. Special care should be taken for the regularization of chiral trace (trace that contains odd number of $\gamma_5$) and the procedure is described in \cite{Xing:2024bpj}. 

The non-zero $\Pi_{ba}(Q)$ is given as follows
\begin{align}
    \Pi_{SS}(Q)&=-4 f_1(Q^2)\\
    \Pi_{PP}(Q)&=4 f_2(Q^2)\\
    \Pi_{PA}(Q)&=4i f_3(Q^2)M Q_{\mu_a}\\
    \Pi_{VV}(Q)&=-8 f_4(Q^2) \delta_{\mu_a\mu_b}^T(Q)\\
    \Pi_{VT}(Q)&=4 f_3(Q^2)M (Q_{\mu_a}\delta_{\nu_a\mu_b}-Q_{\nu_a}\delta_{\mu_a\mu_b})\\
    \Pi_{AP}(Q)&=-4i f_3(Q^2)M Q_{\mu_b}\\
    \Pi_{AA}(Q)&=8f_5(Q^2)\delta_{\mu_a\mu_b}-8f_4(Q^2) Q_{\mu_a} Q_{\mu_b}\\
    \Pi_{TV}(Q)&=-4 f_3(Q^2)M (Q_{\mu_b}\delta_{\nu_b\mu_a}-Q_{\nu_b}\delta_{\mu_b\mu_a})\\
    \Pi_{TT}(Q)&=4f_{6}(Q^2) (\delta_{\mu_a\mu_b}\delta_{\nu_a\nu_b}-\delta_{\mu_a\nu_b}\delta_{\mu_b\nu_a})\nn
               &+8f_4(Q^2) (Q_{\mu_a}Q_{\mu_b}\delta_{\nu_a\nu_b}+Q_{\nu_a}Q_{\nu_b}\delta_{\mu_a\mu_b})\nn
               &-8f_4(Q^2) (Q_{\mu_a}Q_{\nu_b}\delta_{\mu_b\nu_a}+Q_{\mu_b}Q_{\nu_a}\delta_{\mu_a\nu_b})
\end{align}
where (with $\omega=M^2+u\bar{u}Q^2$ and $\bar{u}=1-u$)
\begin{align}
    f_{1}(Q^2)&= \int_0^1 I_2(\omega)-2\omega I_0(\omega)\, du\\
    f_{2}(Q^2)&= \int_0^1 I_2(\omega)-2u\bar{u}Q^2 I_0(\omega)\,du\\
    f_{3}(Q^2)&= \int_0^1 I_0(\omega)\,du\\
    f_{4}(Q^2)&=- \int_0^1 u\bar{u} I_0(\omega)\,du\\
    f_{5}(Q^2)&=- \int_0^1 \omega I_0(\omega)\,du\\
    f_{6}(Q^2)&= \int_0^1 I_2(\omega)+2\omega I_0(\omega)\,du
\end{align}
The regularized expressions for $\mathcal{T}_{cab}$ is lengthy and not presented herein, but it can be reproduced following the regularization procedure in \cite{Xing:2022jtt,Xing:2024bpj}
\bibliography{mainnew}
\end{document}